\documentclass[12pt]{article}
\usepackage{amsmath}
\usepackage[left=1in, right=1in, top=1in, bottom=1in]{geometry}
\usepackage{graphicx}
\usepackage{enumerate}
\usepackage{natbib}
\usepackage{url} 
\usepackage{titling}
\usepackage{hyperref}
\usepackage{setspace}
\usepackage{booktabs}
\usepackage{empheq}
\usepackage{amsfonts}
\usepackage[compact]{titlesec}
\usepackage[labelfont=bf]{caption}
\usepackage{amssymb}
\usepackage{amsthm}
\usepackage{bm}
\usepackage{titletoc}
\usepackage{verbatim}
\usepackage{xcolor}
\usepackage{centernot}
\usepackage{mathtools}
\usepackage{multirow}
\usepackage{fancyhdr}
\usepackage{acronym}
\usepackage{adjustbox}
\usepackage{subcaption}
\usepackage{textcase}
\usepackage{enumitem}
\usepackage{breqn}
\usepackage{xcolor}
\usepackage{authblk}
\usepackage{changepage}

\RequirePackage[labelfont=bf,font=small,skip=3pt]{caption}

\setlist{noitemsep}
\hypersetup{
  colorlinks   = true,
  urlcolor     = blue,
  linkcolor    =black,
  citecolor    = blue
}

\newcommand{\changefont}{%
    \fontsize{6}{11}\selectfont
}

\newcommand{\sgn}{\text{sign}}

\makeatletter
\newcommand*\bigcdot{\mathpalette\bigcdot@{.5}}
\newcommand*\bigcdot@[2]{\mathbin{\vcenter{\hbox{\scalebox{#2}{$\m@th#1\bullet$}}}}}
\makeatother
\DeclareMathOperator*{\argmax}{arg\,max}

\DeclareMathOperator*{\MD}{\mathcal{D}}

\DeclareMathOperator*{\BX}{\mathbf{X}}
\DeclareMathOperator*{\Bx}{\mathbf{x}}

\renewcommand\and{\end{tabular}\kern-\tabcolsep\ and\ \kern-\tabcolsep\begin{tabular}[t]{c}}
\let\origthanks\thanks
\renewcommand\thanks[1]{\begingroup\let\rlap\relax\origthanks{#1}\endgroup}

\titlespacing*{\section} {0pt}{4pt}{4pt}
\titlespacing*{\subsection} {0pt}{3pt}{3pt}

\renewenvironment{abstract}
 {\small
  \begin{center}
  \bfseries \abstractname\vspace{-.5em}\vspace{0pt}
  \end{center}
  \list{}{
    \setlength{\leftmargin}{.5cm}%
    \setlength{\rightmargin}{\leftmargin}%
  }%
  \item\relax}
 {\endlist}

\title{\vspace{-2.5cm} \bf Sequential Re-estimation Learning of Optimal Individualized Treatment Rules Among Ordinal Treatments with Application to Recommended Intervals Between Blood Donations\thanks{This work was supported by the UK Medical Research Council programme MC\_UU\_00002/2 and the Cambridge International Scholarship. The academic coordinating centre for INTERVAL was supported by core funding from: NIHR Blood and Transplant Research Unit in Donor Health and Genomics (NIHR BTRU-2014-10024), UK Medical Research Council (MR/L003120/1), British Heart Foundation (SP/09/002; RG/13/13/30194; RG/18/13/33946) and the NIHR [Cambridge Biomedical Research Centre at the Cambridge University Hospitals NHS Foundation Trust].}}
\date{}

\author[1]{Yuejia Xu\thanks{\leftskip=-0.6cm yuejia.xu@mrc-bsu.cam.ac.uk}}
\author[2,3]{Angela M. Wood}
\author[4,5]{David J. Roberts}
\author[1]{Brian D. M. Tom\thanks{\leftskip=-0.6cm brian.tom@mrc-bsu.cam.ac.uk 
 }}
\affil[1]{\small MRC Biostatistics Unit, University of Cambridge}
\affil[2]{\small Cardiovascular Epidemiology Unit, University of Cambridge}
\affil[3]{\small NIHR Blood and Transplant Research Unit in Donor Health and Genomics}
\affil[4]{\small BRC Haematology Theme and Radcliffe Department of Medicine, University of Oxford}
\affil[5]{\small National Health Service Blood and Transplant}

\begin{document}

\newgeometry{left=0.9in,right=0.9in,bottom=1in}

\maketitle

\def\spacingset#1{\renewcommand{\baselinestretch}%
{#1}\small\normalsize} \spacingset{1}


%
%

\vspace{-2.5em}

\begin{abstract}

Personalized medicine has gained much popularity recently as a way of providing better healthcare by tailoring treatments to suit individuals. Our research, motivated by the UK INTERVAL blood donation trial, focuses on estimating the optimal individualized treatment rule (ITR) in the ordinal treatment-arms setting. Restrictions on minimum lengths between whole blood donations exist to safeguard donor health and quality of blood received. However, the evidence-base for these limits is lacking. Moreover, in England, the blood service is interested in making blood donation both safe and sustainable by integrating multi-marker data from INTERVAL and developing personalized donation strategies. As the three inter-donation interval options in INTERVAL have clear orderings, we propose a sequential re-estimation learning method that effectively incorporates ``treatment" orderings when identifying optimal ITRs. Furthermore, we incorporate variable selection into our method for both linear and nonlinear decision rules to handle situations with (noise) covariates irrelevant for decision-making. Simulations demonstrate its superior performance over existing methods that assume multiple nominal treatments by achieving smaller misclassification rates and larger value functions. Application to a much-in-demand donor subgroup shows that the estimated optimal ITR achieves both the highest utilities and largest proportions of donors assigned to the safest inter-donation interval option in INTERVAL.
\end{abstract}

\noindent%

\restoregeometry

\doublespacing
\section{Introduction}\label{sec:intro}
Responses to a given treatment can vary substantially across patients due to heterogeneity in patient-specific characteristics, such as demographic information, clinical measurements, biological markers, etc. Therefore, a universal strategy that treats all patients with the same treatment may be inadequate and lead to suboptimal treatment decisions. Instead, an individualized clinical decision-making strategy that accounts for such heterogeneity and tailors treatment to individual patients are more desirable in medical practice.

Our research was motivated by the INTERVAL blood donation trial that was embedded in the UK National Blood Service \citep{moore2014interval,angelantonio2017interval}. INTERVAL was the first randomized trial to investigate the effect of different inter-donation intervals on blood supply and donor health. In the UK, the current practice is to allow men to donate blood at most once every 12 weeks and women at most once every 16 weeks, whereas in other countries, the inter-donation interval can be much shorter. For example, in the US, men and women can donate blood every 8 weeks \citep{karp2010frequency}. In INTERVAL, men were randomly assigned to 12-week (standard), 10-week, or 8-week inter-donation intervals (1:1:1), and women to 16-week (standard), 14-week, or 12-week intervals (1:1:1). INTERVAL participants were well-characterized at baseline, enabling the blood service to explore a more personalized donation strategy that could potentially help safeguard donor health and maintain blood supply. For example, it might be the case that some donors should give blood less frequently to maintain adequate iron store levels, whereas other donors are able to give blood more frequently, depending on a number of factors such as age, body mass index, donation history and levels of haemoglobin and iron stores available at the donation session. We therefore aim to integrate multi-marker data from the INTERVAL trial and develop statistical models which can be used to predict the optimal individualized inter-donation interval that tailors to each donor's donation capacity. In this context, the three options of the inter-donation interval from the INTERVAL trial can be viewed as three ordinal ``treatments". 

Most existing methods for estimating the optimal individualized treatment rule (ITR) can handle only two treatments \citep{zhao2012owl,zhang2012dr,tian2014modifiedcovariate,zhou2017rwl,liu2018aol}. Recently, a few approaches have been developed for clinical studies with more than two nominal (unordered) treatments. For example, \citet{tao2017acwl} proposed the adaptive contrast weighted learning (ACWL), which used adaptive contrasts to recast the problem of multiple treatment comparisons to a weighted classification problem. \citet{qi2018dlearn} proposed the multi-category direct learning (D-Learn) approach which compared multiple treatments based on the effect measure of each treatment option. \citet{zhou2019som} developed the sequential outcome-weighted multicategory (SOM) learning method which addressed the multiple treatment comparison problem via a combination of weighted binary classifications. To our knowledge, there is little research on estimation of the optimal ITR among ordinal treatments. However, in many applications, different treatment options exhibit a natural order. 
When methods developed for nominal treatments are applied to ordinal treatments, one may expect to lose some useful information on treatment orderings that can potentially help improve the prediction performance of the estimated ITR. This may result in suboptimal decisions. In the blood donation context, assuming three inter-donation interval options as ordinal instead of nominal can be particularly beneficial for identifying the optimal donation strategy. For example, for a female donor whose true optimal inter-donation interval is 16-week, incorrectly allocating her to the 12-week inter-donation interval might lead to more severe consequences on donor health than to the 14-week one which is closer to the true optimal. 

In this paper, we propose a sequential re-estimation (SR) learning approach to identify the optimal ITR among ordinal treatments for both linear and nonlinear decision rules. SR learning exploits and effectively incorporates information on ordinality of treatment arms and thus avoids unnecessary pairwise comparisons. Specifically, by taking advantage of treatment orderings, we first decompose the optimal ITR estimation problem into a sequence of binary treatment comparison subproblems, including sequential ones that determine whether a more ``intensive" treatment should be given and re-estimation ones that compare two consecutive treatment categories. Existing methods for estimating the optimal treatment rule in the 2-arm case can subsequently be applied to solve each binary subproblem. For example, we employ the augmented outcome-weighted learning (AOL) method proposed by \citet{liu2018aol} whereby optimal binary decision rules can be estimated under the weighted classification framework. We then ensemble multiple binary decisions predicted by binary classifiers and derive the optimal ITR among ordinal treatments based on a decision tree. 

Clinical studies typically collect a large amount of patient information, but some of them may be irrelevant for making treatment decisions and it is usually challenging to acquire \textit{a priori} knowledge on which patient characteristics are truly helpful. Inclusion of unimportant covariates when estimating the optimal ITR may lead to poor model performance and excessively complicated decision rules \citep{gunter2011qualselect,song2015vs}. Therefore, variable selection is vital for deriving the optimal ITR in order to remove covariates that are unnecessary and reduce the complexity of treatment decision rules. In light of this, we further propose variable selection methods for linear and nonlinear decision rules, respectively, which help improve the performance of SR learning in the presence of noise covariates.

The rest of the paper is organized as follows. In Section \ref{sec:method}, we introduce the statistical framework and the main idea of SR learning for estimating the optimal ITR in the ordinal treatment setting. We also develop variable selection techniques under the proposed framework. In Section \ref{sec:simulation}, we conduct extensive simulation studies to evaluate the finite sample performance of SR learning. In Section \ref{sec:application}, we illustrate our proposed method by applying it to the motivating example (INTERVAL trial). We conclude this paper with discussion in Section \ref{sec:discussion}.

\section{Methodology}\label{sec:method}
\subsection{Notations and Statistical Frameworks}\label{subsec:notation}
\noindent We assume that the data are collected from a clinical trial with $n$ subjects and $K$ ordinal treatments ($K\geq3$). Let $A\in \mathcal{A}=\{1,\ldots,K\}$ denote the treatment assignment, $Y\in\mathbb{R}$ be the target outcome, and $\BX=(X_1,\ldots,X_p)^T\in \mathcal{X}$ be a $p$-dimensional covariate (feature) vector. We observe $(Y_i,\BX_i,A_i)$, for $i=1,\ldots,n$, which are independent and identically distributed across $i$. Given the natural ordering of treatments $1,\ldots,K$, the reference arm can be either treatment 1 or $K$ (1 is the least ``intensive" treatment option and $K$ is the most ``intensive" one).  
Without loss of generality, we assume treatment 1 is the reference arm and a larger $Y$ is more desirable in the following discussion. An individualized treatment rule (ITR), denoted by $\MD$, is a map from the feature space, $\mathcal{X}$, to the domain of treatment assignment, $\mathcal{A}$. We assume the propensity score $P(A=a|\BX=\Bx)>0$ with probability 1, $\forall$ $(\Bx,a)\in \mathcal{X}\times \mathcal{A}$. The assumptions of ``consistency" and ``no unmeasured confounders (NUC)" are typically standard and satisfied in the clinical trial setting \citep{robins2000causal}. \citet{qian2011value} introduced the ``value function'' associated with the treatment rule $\mathcal{D}$ as follows: 
\begin{equation}\label{eqn:value}
V(\MD)=E\bigg[\frac{I\big\{A=\MD(\BX)\big\}}{P(A|\BX)}Y\bigg],
\end{equation}
where $E(\bigcdot)$ is the expectation and $I(\bigcdot)$ is the indicator function. $V(\MD)$ can be interpreted as the expected outcome had all subjects in the given population followed the rule $\mathcal{D}$, and we aim to find the optimal ITR, $\MD^*$, that maximizes $V(\MD)$, i.e., $\mathcal{D}^*=\argmax_{\MD}E\big[{I\big\{A=\MD(\BX)\big\}Y}/{P(A|\BX)}\big].$ 
\noindent There are many well-established methods for estimating $\mathcal{D}^*$ in the two-arm setting \citep{zhang2012dr,zhao2012owl,tian2014modifiedcovariate,zhou2017rwl,liu2018aol}. However, extension to the ordinal setting is nontrivial. 

\subsection{Sequential Re-estimation Learning for Ordinal Treatments}\label{sec:SR}
In this section, we propose a method that takes advantage of the ordering information on treatment arms and estimates the optimal ITR among ordinal treatments. Specifically, we decompose the problem of estimating the optimal ITR for ordinal treatments into multiple subproblems of binary treatment comparisons, which subsequently can be solved using existing methods developed for the binary treatment setting (details on training binary classifiers will be discussed later in Section \ref{subsec:binary}). We refer to the proposed method as sequential re-estimation (SR) learning. 

\subsubsection{Learning}
The model learning process consists of $K-1$ ``sequential" and $K-2$ ``re-estimation" steps.
\paragraph{Sequential step (S-step)}
Each sequential step trains a binary classifier $S_k$ of treatment $\{k\}$ vs. $\{k+1,\ldots,K\}$, $k=1,\ldots,K-1$. For $k=1$, all individuals are included when training $S_1$, and for $k=2,\ldots,K-1$, we only include individuals whose observed treatments $A_i$ do not belong to $\{1,\ldots,k-1\}$ and whose optimal treatments are not estimated as $j$ in the $j^{\text{th}}$ sequential step for all $j<k$ \citep{zhou2019som}, i.e., those who satisfy $I\big(A_i \notin\{1,\ldots,k-1\},\widehat{S}_1(\mathbf{X}_i)\neq 1, \ldots,\widehat{S}_{k-1}(\mathbf{X}_i)\neq k-1\big)=1$. Intuitively, the classifier $S_k$ determines whether or not treatment options that are more ``intensive" than treatment $k$ can lead to more desirable outcomes than $k$. 
\paragraph{Re-estimation step (R-step)}
Each re-estimation step trains a binary classifier $R_k$ of treatment $\{k\}$ vs. $\{k+1\}$ 
using individuals whose observed treatments $A_i$ are either $k$ or $k+1$ and whose optimal treatments are estimated as $k$ in the $k^{\text{th}}$ sequential step, i.e., those satisfying $I\big(A_i\in\{k ,k+1\},\widehat{S}_k(\mathbf{X}_i)=k\big)=1$, for $k=1,\ldots,K-2$. 

The purpose of training $K-2$ R-step classifiers, in addition to S-step classifiers, is to estimate decision boundaries between two ``consecutive" treatment categories $k$ and $k+1$ more accurately by using the data from a more ``refined" population. Moreover, it allows subjects who have been assigned to receive ``conservative" treatments in S-steps to be re-considered for whether a treatment ``step-up'' may benefit them. We provide justifications for the importance of R-steps in the supplementary material S.1, where we plot estimated boundaries by S- and R-steps, and present classification results by category from simulated examples to highlight substantial improvement from the inclusion of R-steps.



We note that the number of eligible subjects who are included in training R-step classifiers is in general smaller than that of subjects who are included in training S-step classifiers. For example, information on all $n$ subjects are used when training the first sequential classifier $S_1$ which determines whether or not a person should receive a more ``intensive" treatment than the reference arm (treatment 1), whereas when training the first re-estimation classifier $R_1$ between treatment 1 and 2, we only use a subset of the data from subjects whose predicted optimal treatments based on $S_1$ are 1 and observed treatments are either 1 or 2. This is closely aligned with how clinical decisions are made in practice. The decision on whether or not a patient should change to a more ``intensive'' intervention (compared to the reference/current practice) is made first before deciding on the actual treatment option to be given. The proposed SR learning trains the S-step classifiers with larger datasets and thus clinicians can be more confident when deciding whether or not a more ``intensive" treatment should be administered.

\subsubsection{Ensembling} 
\begin{figure}[]
\centering
\begin{subfigure}{.4\textwidth}
  \centering
  \includegraphics[width=\linewidth,height=5cm,keepaspectratio]{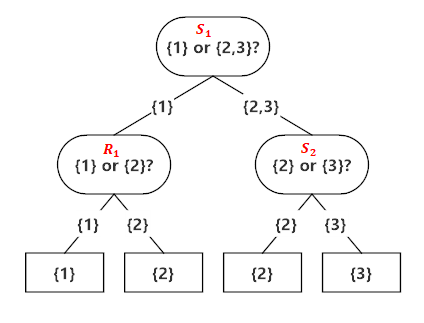}
  \caption{$K=3$}
  \label{fig:sub1}
\end{subfigure}%
\begin{subfigure}{.55\textwidth}
  \centering
  \includegraphics[width=\linewidth,height=7.5cm,keepaspectratio]{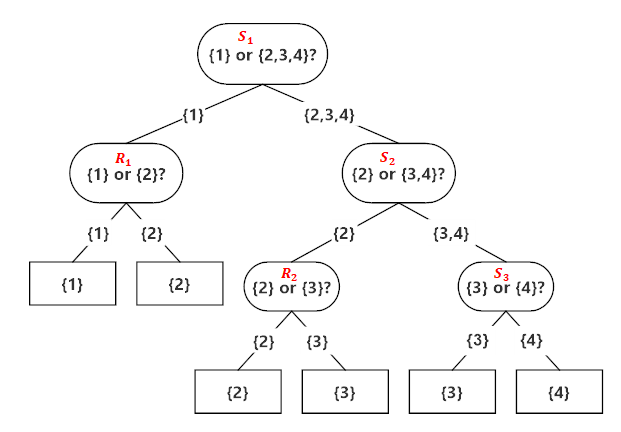}
  \caption{$K=4$}
  \label{fig:sub2}
\end{subfigure}
\caption{Illustration of how multiple binary decisions are aggregated for ordinal prediction.}\label{fig:tree}
\end{figure}

For a given subject with covariates $\mathbf{x}$, we can estimate his/her optimal treatment based on $\widehat{S}_k(\mathbf{x})$ and $\widehat{R}_k(\mathbf{x})$. We demonstrate the idea of aggregating multiple binary treatment selection decisions and predicting the optimal treatment among $K$ ordinal options by tree diagrams in \autoref{fig:tree} ((a) for $K=3$ and (b) for $K=4$). 

For example, when $K=3$, the optimal treatment for a subject with covariates $\mathbf{x}$ is $I\big(\widehat{S}_1(\mathbf{x})=1\big)\widehat{R}_1(\mathbf{x})+I\big(\widehat{S}_1(\mathbf{x})\neq 1)\widehat{S}_2(\mathbf{x})$, implying that if the first sequential step favors \{1\} over \{2,3\}, then the final decision is determined by the re-estimation step where \{1\} and \{2\} are compared. Otherwise, the decision depends on the second sequential step. When $K=4$, the optimal treatment is given by $I\big(\widehat{S}_1(\mathbf{x})=1\big)\widehat{R}_1(\mathbf{x})+I(\widehat{S}_1(\mathbf{x}) \neq 1,\widehat{S}_2(\mathbf{x})=2)\widehat{R}_2(\mathbf{x})+I(\widehat{S}_1(\mathbf{x}) \neq 1,\widehat{S}_2(\mathbf{x})\neq 2)\widehat{S}_3(\mathbf{x})$. The idea follows similarly for $K>4$, where the optimal decision is:
\begin{equation}
\begin{split}
 \widehat{\mathcal{D}^*}(\Bx)=&I\big(\widehat{S}_1(\mathbf{x})=1\big)\widehat{R}_1(\mathbf{x})+I\big(\widehat{S}_1(\mathbf{x})\neq 1,\widehat{S}_2(\mathbf{x})=2\big)\widehat{R}_2(\mathbf{x})+\ldots\\
 &+I\big(\widehat{S}_1(\mathbf{x})\neq 1,\ldots,\widehat{S}_{K-3}(\mathbf{x})\neq K-3, \widehat{S}_{K-2}(\mathbf{x})=K-2\big)\widehat{R}_{K-2}(\mathbf{x})\\&+I\big(\widehat{S}_1(\mathbf{x})\neq 1,\ldots,\widehat{S}_{K-2}(\mathbf{x})\neq K-2\big)\widehat{S}_{K-1}(\mathbf{x}).
\end{split}
\end{equation}

\subsection{Training Binary Classifiers $S_k$ and $R_k$}\label{subsec:binary} 
In principle, all methods developed for estimating the optimal ITR with two treatment options are applicable for learning binary decision rules in S- and R-steps. In practice, due to the reduced sample sizes in intermediate steps, we recommend using methods with high convergence rate and desirable small sample performance for binary comparisons \citep{zhou2017rwl,liu2018aol}. We follow \citet{zhou2019som} and adopt in our implementation the augmented outcome-weighted learning (AOL) method proposed by \citet{liu2018aol} for solving binary subproblems and training binary classifiers $S_k$ and $R_k$ in SR learning. In the following discussion, we take $S_k$ as an example and focus on deriving the training process for $S_k$. $R_k$ can be trained in a similar way. 

For notational clarity, we denote the new treatment label for training $S_k$ by $A^{S_k}$, the number of eligible subjects included in training $S_k$ by $n^{S_k}$, the propensity score associated with the binary treatment label by $P^{S_k}(A^{S_k}|\BX)$, and an ITR between $\{k\}$ vs. $\{k+1,\ldots,K\}$ in step $S_k$ by $\MD^{S_k}$, respectively, for $k=1,\ldots,K-1$. We note that $P^{S_k}(A^{S_k}|\BX)$ can be estimated via logistic regression. Without loss of generality, we assume $A^{S_k}\in \{-1,1\}$. Similar to (\ref{eqn:value}), the value function associated with $\mathcal{D}^{S_k}$ for the binary comparison in step $S_k$ can be written as \begin{equation}\label{eqn:value_binary}
V(\mathcal{D}^{S_k})=E\bigg[\frac{I\big\{A^{S_k}=\MD^{S_k}(\BX)\big\}}{P^{S_k}(A^{S_k}|\BX)}Y\bigg].
\end{equation}
The optimal ITR can be estimated by maximizing an empirical version of (\ref{eqn:value_binary}) based on the observed data, that is, $\frac{1}{n^{S_k}}\sum_{i=1}^{n^{S_k}}{I\big\{A^{S_k}_i =\MD^{S_k}(\BX_i)\big\}Y_i}/{P^{S_k}(A^{S_k}_i|\BX_i)}$,
or equivalently, minimizing the empirical weighted misclassification rate, $\frac{1}{n^{S_k}}\sum_{i=1}^{n^{S_k}}{I\big\{A^{S_k}_i \neq \MD^{S_k}(\BX_i)\big\}Y_i}/{P^{S_k}(A^{S_k}_i|\BX_i)}.$
\citet{liu2018aol} showed that this minimization problem can be modified to improve efficiency by minimizing the following expression:
\begin{equation}\label{eqn:aol_1}
\frac{1}{n^{S_k}}\sum_{i=1}^{n^{S_k}}\frac{I\big\{A^{S_k}_i\sgn(e_i) \neq \MD^{S_k}(\BX_i)\big\}|e_i|}{P^{S_k}(A^{S_k}_i|\BX_i)},
\end{equation}
where $e_i=Y_i-\widehat m(\BX_i)$, and $\widehat m(\BX_i)$ is the fitted value from the model that regresses $Y_i$ on $\BX_i$. As suggested by \citet{zhao2012owl}, any ITR $\MD^{S_k}(\BX_i)$ can be represented as $\MD^{S_k}(\BX_i)=\sgn(f^{S_k}(\BX_i))$, for some measurable decision function $f^{S_k}(\BX_i)$. Therefore, (\ref{eqn:aol_1}) is equivalent to 
\begin{equation} \label{eqn:aol_2}
\frac{1}{n^{S_k}}\sum_{i=1}^{n^{S_k}}\frac{I\big\{A^{S_k}_i\sgn(e_i)f^{S_k}(\BX_i)<0 \big\}|e_i|}{P^{S_k}(A^{S_k}_i|\BX_i)}.
\end{equation}
The zero-one loss function is discontinuous and nonconvex and thus minimizing (\ref{eqn:aol_2}) is NP-hard. This problem can be addressed by replacing the zero-one loss with the hinge loss and estimating the optimal decision function by solving a weighted support vector machine (SVM) problem \citep{zhao2012owl}. The estimated optimal decision function ${\widehat{f^*}}^{S_k}(\BX)$ minimizes
\begin{equation}\label{eqn:aol_3}
 \frac{1}{n^{S_k}}\sum_{i=1}^{n^{S_k}}\frac{\{1-A^{S_k}_i\sgn(e_i)f^{S_k}(X_i)\}^{+}|e_i|}{P^{S_k}(A^{S_k}_i|\BX_i)}+\lambda\Vert f^{S_k} \Vert^2,
\end{equation}
where $(\bigcdot)^+=\text{max}(0,\bigcdot)$. An $l_2$ regularization term $\lambda\Vert f^{S_k} \Vert^2$ is included in the objective function (\ref{eqn:aol_3}) to avoid overfitting, where $\lambda$ is the tuning parameter \citep{zhao2012owl}. 

 We note that the decision function $f^{S_k}$ can either be linear or nonlinear. For linear decision functions, $f^{S_k}(\BX)=\beta_0+{\BX}^T\bm{\beta}$. For nonlinear decision functions, $f^{S_k}(\BX)$ can be written as $f^{S_k}(\BX)=\beta_0+h(\BX)$, where $h(\BX) \in \mathcal{H}_{\mathcal{K}}$ and $\mathcal{H}_{\mathcal{K}}$ is a reproducing kernel Hilbert space (RKHS) associated with the kernel function ${\mathcal{K}}(\bigcdot,\bigcdot)$ (maps from $\mathcal{X} \times \mathcal{X}$ to $\mathbb{R}$). For example, the Gaussian radius basis function (RBF) kernel is a commonly used nonlinear kernel and can be written as ${\mathcal{K}}_{\sigma}(u,v)=\text{exp}\big(-{\Vert u-v \Vert} ^2/2 \sigma^2\big)$, where $\sigma>0$ is the bandwidth parameter that determines how far the influence of a data point reaches. We provide details on solving (\ref{eqn:aol_3}) and estimating the optimal decision function in the supplementary material S.2. The optimal ITR can subsequently be estimated as ${\widehat{\MD^*}}^{S_k}(\BX)=\sgn({\widehat{f^*}}^{S_k}(\BX))$.

\subsection{Variable Selection}\label{subsec:vs}
With rapid technological advancement in collecting individual-level information, an increasing number of clinical and biological covariates can be measured and are available in clinical studies. However, some information may be unnecessary and substantial resources can be saved by measuring only relevant covariates. In addition, in the presence of high-dimensional covariates, noise variables may ``pollute" and impair model performance (increase computational time, affect convergence, and decrease generalization ability and prediction accuracy) due to the limited sample sizes in clinical studies - a phenomenon commonly referred to as ``the curse of dimensionality". Interpretability is also a major concern in this case as models that are fitted with many covariates can be hard to interpret. In the context of estimating the optimal ITR, only covariates that interact qualitatively with the treatment are of clinical importance for decision-making; and thus it is crucial to identify key covariates that have impact on the optimal treatment decisions through variable selection, and to derive simple and practically implementable decision rules \citep{gunter2011qualselect,song2015vs}. 


When the dimension of the covariate space is relatively high, we follow the SR learning framework introduced in Section \ref{sec:SR} and incorporate the variable selection feature into the training of each binary classifier $S_k$ and $R_k$ to select a subset of covariates and improve the discriminative ability of each classifier. Selecting important covariates for each classifier independently also makes sense in practice, since it is reasonable to think that boundaries between ``consecutive" treatment categories may depend on different covariates, and performing variable selection independently for each classifier allows for this flexibility. As before, we take $S_k$ as an example. $R_k$ can be trained in a similar manner.

\subsubsection{Linear Decision Function}\label{subsec:vs_linear}
In Section \ref{subsec:binary}, the objective function (\ref{eqn:aol_3}) for training binary classifiers $S_k$  includes the $l_2$ penalty term $\lambda\Vert f^{S_k} \Vert^2$. It is well-known that $l_2$ penalty shrinks coefficients towards zero, but does not perform variable selection nor lead to sparse solutions. In contrast, $l_1 $ penalty allows some coefficients to be exactly zero when $\lambda$ is sufficiently large and inherently performs variable selection \citep{tibshirani1996lasso}. Analogous to  the 1-norm SVM formulation \citep{zhu20031normsvm,mangasarian20061norm}, we replace the $l_2$ penalty term $\lambda\Vert f^{S_k} \Vert^2$ in (\ref{eqn:aol_3}) with the $l_1$ penalty $\lambda\Vert f^{S_k} \Vert$ to incorporate the variable selection feature into the AOL framework. The optimal decision function can be estimated by minimizing 
\begin{equation}\label{eqn:aol_vs_linear}
 \frac{1}{n^{S_k}}\sum_{i=1}^{n^{S_k}}\frac{\{1-A^{S_k}_i\sgn(e_i)f^{S_k}(X_i)\}^{+}|e_i|}{P^{S_k}(A^{S_k}_i|\BX_i)}+\lambda\Vert f^{S_k} \Vert,
\end{equation}
where $\lambda$ is the tuning parameter that controls the trade-off between the misclassification error and the complexity of the estimated decision function. For linear decision functions, $f^{S_k}(\BX)=\beta_0+{\BX}^T\bm{\beta}$, and we formulate (\ref{eqn:aol_vs_linear}) as a linear programming problem that can be solved efficiently using the simplex algorithm (details are provided in the supplementary material S.3.1). 

\subsubsection{Nonlinear Decision Function}\label{subsec:vs_nonlinear}
We can apply similar techniques discussed in Section \ref{subsec:vs_linear} to nonlinear decision functions. However, doing this results in a penalized kernel space instead of a penalized input space, i.e., instead of reducing the number of input space features as in the linear case, imposing $l_1$ penalty in the nonlinear case reduces the number of kernel functions used to generate the nonlinear classifier (dimensionality of the higher dimensional transformed space) \citep{mangasarian2007nonlinearsvm,dasgupta2019vs}. In the SVM literature, \citet{mangasarian2007nonlinearsvm} addressed this problem by introducing a diagonal matrix $\mathbf{E}$ with diagonal entries being either zero or one, where zeros correspond to eliminated features and ones correspond to features utilized for constructing nonlinear decision rules. They then solved a mixed-integer nonlinear programming problem by alternating between solving a linear programming problem and updating the diagonal elements of $\mathbf{E}$. One major issue with this approach is that the mixed-integer programming problem is nonconvex \citep{nguyen2010vs}. Therefore, the optimization procedure under the formulation proposed by \citet{mangasarian2007nonlinearsvm} can be computationally demanding, very sensitive to starting values, and may be stuck at local optima rather than converging to a solution that is globally optimal \citep{boyd2004convexoptimization}. 

In practice, picking a decent starting value for a moderately high-dimensional vector can be challenging. For this reason, instead of seeking a good starting value for the nonconvex optimization problem, we propose a two-stage procedure to select important covariates and learn nonlinear classification rules when the number of covariates collected in clinical studies is relatively large. The basic idea is to first identify a subset of informative covariates that predict treatment effect heterogeneity with first-order effects $X_j$ or second-order effects $X_jX_k$ ($1 \leq j\leq k \leq p$; quadratic terms when $j=k$ and two-way interactions when $j\neq k$), and then train binary classifiers with nonlinear decision functions using selected covariates. This two-stage procedure is similar to the variable selection approach proposed by \citet{bi2003dimreductionsvm}, but methods that we employ for covariate screening in the first stage and for nonlinear classification in the second stage are different from those used by \citet{bi2003dimreductionsvm}. 

Specifically, in the first stage, we ``pre-screen" covariates using the ``stepwise conditional likelihood variable selection for discriminant analysis (SODA)" method proposed by \citet{li2018soda}. SODA was developed for solving high-dimensional classification problems under the logistic regression framework. It performed variable selection for both first-order and second-order terms in a robust and efficient manner through a stepwise procedure and was shown to enjoy superior performance in terms of variable selection accuracy, classification accuracy, and robustness to non-normality of covariate distributions. We refer readers to the original publication for details \citep{li2018soda}. In our context, we apply SODA to train a logistic regression classifier with the class label $A^{S_k}\sgn(e)/2+1/2$ and covariates $(\BX,\BX \otimes \BX)$, where $\BX \otimes \BX$ represents all second-order terms of $\BX$ (quadratic terms and two-way interaction terms). We retain covariates that are part of any selected monomials.  

In the second stage, we estimate the nonlinear decision function under the framework similar to that proposed by \citet{mangasarian2007nonlinearsvm}, except that we fix the diagonal matrix $\mathbf{E}$ based on variable selection results from the first stage, such that the optimization problem becomes a linear programming one that is convex and can be easily solved using standard linear programming packages. We provide details on the formulation of the optimization problem in the supplementary material S.3.2. 

We note that even though the pre-screening stage focuses only on first-order and second-order terms and may fail to capture more complicated nonlinear structures, in most medical applications, covariates that interact with the optimal treatment decision in a more complex fashion lead to harder-to-interpret rules and are less useful \citep{qiu2018lineartrt}. In addition, the relationship between covariates selected in the first stage and the final optimal decision rule is re-evaluated without the second-order restriction by training the nonlinear classifier using the Gaussian RBF kernel in the second stage. Therefore, SODA should be sufficient for the purpose of covariate screening. We also demonstrate through simulation studies in the next section that in most cases, this pre-screening method suffices for identifying covariates that inform decision-making, and it performs well even when true underlying decision boundaries involve nonlinear terms other than the second-order ones.

We also comment that although variable selection in the first stage and estimation of nonlinear classifiers in the second stage seem to be performed independently, we consider the proposed two-stage procedure as a ``wrapper" method (i.e., variables are selected based on their usefulness in optimizing the classification performance) rather than a ``filter" method (i.e., variables are filtered independently of the classification algorithm) for variable selection \citep{kohavi1997wrapper,guyon2003vs}. This is because we take into consideration the effects of selected covariates on classification performance by wrapping variable selection around the logistic regression framework. Indeed, the two stages are tightly coupled since they both seek to optimize classification performance, and the inclusion of the first stage adds ``robustness" to boundary estimations in the second stage despite the fact that different classification algorithms are implemented for each stage. As pointed out by \citet{bi2003dimreductionsvm}, another major advantage of this type of two-stage procedure for variable selection is that the relevance of covariates in classification can be assessed in a computationally cheaper way compared to the ``embedded" method that directly wraps variable selection around a nonlinear SVM classifier.

\section{Simulation Studies}\label{sec:simulation}
We conduct simulation studies to assess the finite sample performance of the proposed sequential re-estimation learning method in both low-dimensional (without noise covariates and all covariates are informative of optimal treatment decisions) and moderate-dimensional (with noise covariates) settings. 

\subsection{Simulation Design and Evaluation Criteria}\label{subsec:sim_setting}

In each simulation setting, covariates $X_1,X_2,\ldots,X_p$ are independently generated from the uniform distribution $\mathcal{U}\{-1,1\}$, and treatment $A$ is sampled uniformly from ${1,2,\ldots,K}$ such that $P(A=a|\BX=\Bx)=1/K$ for all $\Bx\in\mathcal{X}$ and $a \in \mathcal{A}$. Similar to \citet{chen2016dosefindingowl}, we assume the outcome $Y$ is normally distributed with mean $\mu(\BX)-\varphi\big\{A,\MD^*(\BX)\big\}$ and variance 1, where $\mu(\BX)$ is the main effect of covariates on the outcome and $\varphi\big\{A,\MD^*(\BX)\big\}$ represents the loss in outcome when the assigned treatment is nonoptimal. For each setting, we consider two training sample sizes: $n=400$ and $n=800$, and repeat the simulation 500 times. We examine 10 settings, which cover scenarios with different numbers of ordinal treatment options, different loss functions for receiving nonoptimal treatments, and a broad set of decision boundaries that reflect ordinality, some of which have rarely been explored previously. We repeat each simulation 500 times. 

Settings 1-6 mimic situations where decision boundaries between two ``consecutive" treatment categories are parallel to each other and are determined by a combination of $X_1$, $X_2$, $X_3$, $X_4$, and $X_5$. Decision boundaries are linear in settings 1 to 5 and nonlinear in setting 6. Similar to dose-finding problems \citep{chen2016dosefindingowl}, when treatments are ordinal, we would expect that incorrectly allocating a patient to a treatment that is closer to his/her true optimal leads to a smaller loss in the outcome. Therefore, we consider two types of losses: the absolute loss, $|A-\MD^*(\BX)|$, in settings 1-5, and the quadratic loss, $\{A-\MD^*(\BX)\}^2$, in setting 6. In addition, we vary intercepts of decision functions in settings 1-4 (e.g. the proportion of subjects with $\MD^*(\BX)=1$ is about 10\% in setting 2, but greater than 60\% in setting 3) to examine the robustness of the proposed method to different distributions of the true optimal ITR, $\MD^*(\BX)$, across treatments $1,\ldots,K$. Details on simulation designs of settings 1-6 are provided in the supplementary material Section S.4.1 (Table S4).

 Settings 7-10 correspond to situations where decision functions are nonparallel and boundaries between ``consecutive" treatment categories are no longer shifts of each other. To facilitate the visualization of these settings, we take $p=2$ and assume only $X_1$ and $X_2$ inform the optimal treatment choice. In setting 7, the two decision boundaries are quarter circle and parabola. For settings 8 and 10, we consider circle boundaries with expanding radius ($K=3$ in setting 8 and $K=4$ in setting 10). Setting 9 examines the case where one decision boundary is smooth (ellipse) and the other is nonsmooth (square). True optimal ITRs and decision boundaries in settings 7-10 are illustrated in \autoref{fig:nonparallel_setting} with corresponding expressions provided in the supplementary material S.4.2.
 
 \begin{figure}[h]
\centering
\includegraphics[width=0.9\textwidth]{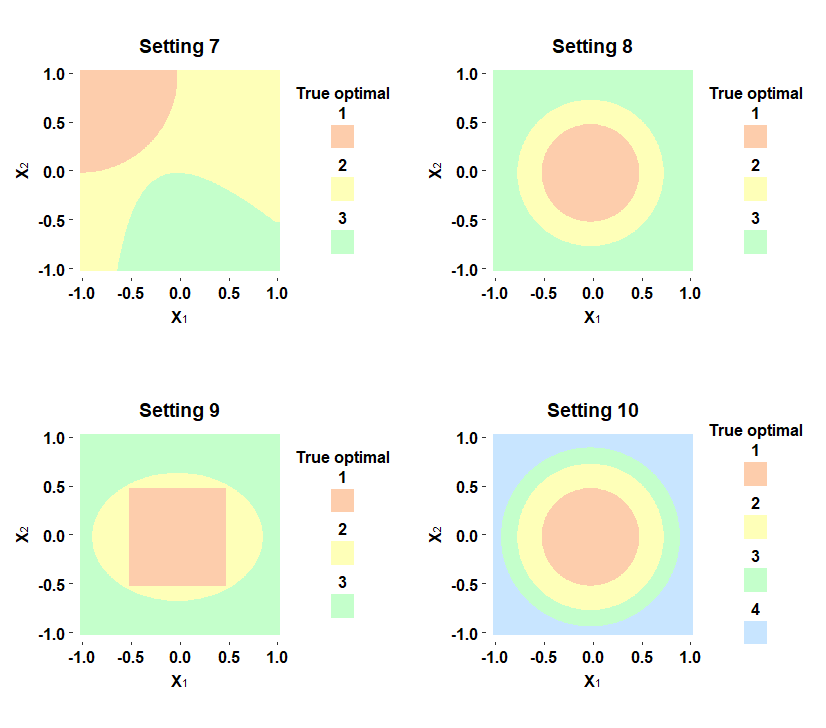}
\caption{Visualization of true underlying decision boundaries and optimal ITRs in simulation settings with nonparallel boundaries (7-10).}\label{fig:nonparallel_setting}
\end{figure}

For settings with parallel boundaries (1-6), we apply SR learning using both linear (SR-Linear) and Gaussian (SR-Gaussian) kernels to train binary classifiers $S_k$ and $R_k$, while for settings with nonparallel boundaries (7-10), we use the more flexible Gaussian kernel to better train classifiers. 
In all settings, we compare SR learning with existing methods that can be used to estimate the optimal ITR in multi-arm trials and incorporate variable selection features, including the multi-category direct learning method (D-Learn) proposed by \citet{qi2018dlearn}, the $l_1$-penalized least squares method (PLS) developed by \citet{qian2011value}, the adaptive contrast weighted learning method introduced by \citet{tao2017acwl} with the minimum contrasts (ACWL-C1) and the maximum contrasts (ACWL-C2). In particular, for D-Learn, we estimate linear decision rules with the least absolute shrinkage and selection operator (LASSO) when true boundaries are linear; and estimate nonlinear decision rules with the component selection and smoothing operator (COSSO) when true boundaries are nonlinear \citep{qi2018dlearn}. For PLS, we use the basis function set $(1,\mathbf{X},A,\mathbf{X}A)$ when true underlying boundaries are linear and basis $(1,\mathbf{X},{\mathbf{X}}^2,A,\mathbf{X}A,{\mathbf{X}}^2A)$ when true boundaries are nonlinear. All tuning parameters are selected by 5-fold cross-validation.

To assess the performance of proposed variable selection methods in situations with noise covariates, we employ the same data-generating process as before and repeat simulations in all settings with $p$ now set to be 50. That means, there are 45 covariates that are unrelated to decision functions for settings 1-6, and 48 irrelevant covariates for settings 7-10. In these moderate-dimensional scenarios, we apply SR learning both with and without variable selection using the kernel that reflects the true boundary types. We still compare our proposed method with D-Learn, PLS, ACWL-C1, and ACWL-C2, since all these methods inherently perform variable selection and thus are directly applicable to cases with moderate-dimensional covariates.

We evaluate the performance of each method on a large independent testing dataset of size 10000 using two criteria: (1) the misclassification rate of the estimated optimal ITR compared to the true optimal ITR, $\mathbb{P}_{n_{\text{test}}}I\{\widehat{\mathcal{D}^*}(\BX)\neq {\mathcal{D}^*}(\BX)\}$, where $\mathbb{P}_{n_{\text{test}}}$ denotes the empirical average over the testing dataset, and (2) the estimated value under $\widehat{\mathcal{D}^*}(\mathbf{X})$ on the testing dataset, $\widehat{V}(\widehat{\mathcal{D}^*})=\mathbb{P}_{n_{\text{test}}}[YI(A=\widehat{\mathcal{D}^*}(\BX))/P(A|\BX)]/\mathbb{P}_{n_{\text{test}}}[I(A=\widehat{\mathcal{D}^*}(\BX))/P(A|\BX)]$. In the simulation settings that we examine, $P(A|\BX)$ is constant. Therefore, estimated value can be interpreted as the average outcome of testing samples whose observed treatments are the same as the estimated optimal ones. 

\subsection{Simulation Results} 
\subsubsection{Low-dimensional $\mathbf{X}$ (Without Noise Covariates)}
In this section, we show results from simulation studies when all components of $\mathbf{X}$ play a role in determining the optimal ITR.
Empirical misclassification rates and values corresponding to parallel settings 1-6 in the low-dimensional case are presented in \autoref{tab:low_dim_parallel} and Figure S2 (supplementary material S.4.3). 
\begin{table}[!htb]
\centering
\caption{Simulation results based on 500 replicates: mean (sd) of misclassification rates and value functions for settings with parallel boundaries (1-6) and no noise covariates. The smallest misclassification rate and largest value function for each setting are in bold.}\label{tab:low_dim_parallel}
\begin{adjustbox}{width=0.75\textwidth,center}
\begin{tabular}{clccccccccc}
\toprule[0.25ex]
 \rule{0pt}{12pt}
          &  & & \multicolumn{3}{c}{$n=400$} && \multicolumn{3}{r}{$n=800$}                                                        \\ \cline{4-6}  \cline{9-11} \rule{0pt}{14pt}
 Setting        & Method        & & Misclassification &  & Value &&& Misclassification & & Value   \\ \hline
 \rule{0pt}{12pt}
\multirow{6}{*}{1} 
			 & D-Learn             &&  0.18 (0.04) 
  && 4.32 (0.17)&&& 0.12 (0.03)&& 4.53 (0.13)\\
                  & PLS            &&  0.11	(0.03)
  && 4.58 (0.11)&&& 0.08 (0.02)&& 4.70 (0.08)\\
                  & ACWL-C1         &&  0.31 (0.02)
  && 3.80 (0.09)&&& 0.28 (0.02)&& 3.91 (0.06)\\
                  & ACWL-C2         &&  0.31 (0.02)
  && 3.79 (0.08)&&& 0.30 (0.02)&& 3.85 (0.08)\\
                  & SR-Linear         && \textbf{0.03 (0.01)}
  && \textbf{4.85 (0.04)}&&& \textbf{0.02 (0.01)}&& \textbf{4.91 (0.03)}\\
                  & SR-Gaussian    &&  0.07 (0.02)
  && 4.72 (0.08)&&& 0.04 (0.01)&& 4.82 (0.05)\\ \hline
 \rule{0pt}{12pt}
 \multirow{6}{*}{2} 
			 & D-Learn     &&  0.17 (0.03)
  && 4.34 (0.13)&&& 0.15 (0.02)&& 4.44 (0.10)\\
                  & PLS      &&  0.14 (0.02)  && 4.46 (0.09)&&& 0.13 (0.02)&& 4.50 (0.07)\\
                  & ACWL-C1    && 0.25 (0.02)  && 3.94 (0.09)&&& 0.22 (0.01)&& 4.08 (0.06)\\
                  & ACWL-C2    && 0.25 (0.02)  && 3.94 (0.09)&&& 0.23 (0.01)&& 4.06 (0.07)\\
                  & SR-Linear    &&  \textbf{0.04 (0.01)}  && \textbf{4.83 (0.06)}&&& \textbf{0.02 (0.01)}&& \textbf{4.91 (0.03)}\\
                  & SR-Gaussian    &&  0.06 (0.01) && 4.75 (0.06)&&& 0.04 (0.01)&& 4.84 (0.04)\\ \hline
 \rule{0pt}{12pt}
 \multirow{6}{*}{3} 
			 & D-Learn     && 0.17 (0.03)  && 4.33 (0.12)&&& 0.15 (0.03)&& 4.41 (0.11)\\
                  & PLS      && 0.14 (0.02)  && 4.44 (0.10)&&& 0.13 (0.02)&& 4.48 (0.08)\\
                  & ACWL-C1    && 0.25 (0.02)  && 3.96 (0.08)&&& 0.23 (0.01)&& 4.09 (0.06)\\
                  & ACWL-C2    && 0.25 (0.02)  && 3.97 (0.08)&&& 0.23 (0.01)&& 4.08 (0.06)\\
                  & SR-Linear    && \textbf{0.04 (0.01)}  && \textbf{4.82 (0.06)}&&& \textbf{0.02 (0.01)}&& \textbf{4.89 (0.02)}\\
                  & SR-Gaussian    && 0.06 (0.02)  && 4.72 (0.08)&&& 0.04 (0.01)&& 4.80 (0.05)\\ \hline
 \rule{0pt}{12pt}
 \multirow{6}{*}{4} 
			 & D-Learn     && 0.15 (0.03)  && 4.42 (0.13)&&& 0.11 (0.02)&& 4.56 (0.10)\\
                  & PLS      && 0.11 (0.02)  && 4.58 (0.09)&&& 0.08 (0.02)&& 4.68 (0.07)\\
                  & ACWL-C1    && 0.30 (0.02)  && 3.67 (0.09)&&& 0.28 (0.01)&& 3.77 (0.08)\\
                  & ACWL-C2    && 0.30 (0.02)  && 3.65 (0.11)&&& 0.28 (0.01)&& 3.71 (0.09)\\
                  & SR-Linear    && \textbf{0.04 (0.01)}  && \textbf{4.85 (0.06)}&&& \textbf{0.02 (0.01)}&& \textbf{4.92 (0.03)}\\
                  & SR-Gaussian    && 0.07 (0.02)  && 4.71 (0.08)&&& 0.05 (0.01)&& 4.82 (0.05)\\ \hline
 \rule{0pt}{12pt}
 \multirow{6}{*}{5} 
			 & D-Learn     && 0.26 (0.05)  && 3.99 (0.20)&&& 0.19 (0.04)&& 4.27 (0.14)\\
                  & PLS      && 0.21 (0.04)  && 4.18 (0.15)&&& 0.15 (0.03)&& 4.40 (0.12)\\
                  & ACWL-C1    && 0.45 (0.02)  && 2.92 (0.16)&&& 0.42 (0.02)&& 3.13 (0.12)\\
                  & ACWL-C2    && 0.43 (0.02)  && 3.00 (0.14)&&& 0.42 (0.02)&& 3.05 (0.14)\\
                  & SR-Linear    && \textbf{0.07 (0.02)}  && \textbf{4.71 (0.10)}&&& \textbf{0.03 (0.01)}&& \textbf{4.87 (0.04)}\\
                  & SR-Gaussian    && 0.13 (0.03)  && 4.45 (0.15)&&& 0.08 (0.02)&& 4.68 (0.08)\\ \hline
 \rule{0pt}{12pt}
 \multirow{6}{*}{6} 
			 & D-Learn     && 0.32 (0.06)  && 5.14 (0.27)&&& 0.28 (0.06)&& 5.30 (0.25)\\
                  & PLS      && 0.23 (0.01)  && 5.54 (0.06)&&& 0.21 (0.01)&& 5.60 (0.04)\\
                  & ACWL-C1    && 0.38 (0.03)  && 4.63 (0.22)&&& 0.35 (0.03)&& 4.83 (0.18)\\
                  & ACWL-C2    && 0.43 (0.03)  && 4.50 (0.23)&&& 0.41 (0.03)&& 4.72 (0.16)\\
                  & SR-Linear    && 0.49 (0.04)  && 4.52 (0.12)&&& 0.49 (0.03)&& 4.56 (0.09)\\
                  & SR-Gaussian    && \textbf{0.16 (0.02)}  && \textbf{5.89 (0.10)}&&& \textbf{0.11 (0.01)}&& \textbf{6.11 (0.06)}\\ 

\bottomrule[0.25ex]
\end{tabular}%
\end{adjustbox}
\end{table}For all these settings, our method performs the best in that it leads to the smallest misclassification rate and the largest value. Among the other competing methods, PLS seems to be slightly better than D-Learn, and much better than tree-based methods ACWL-C1 and ACWL-C2. In settings 1-5, decision boundaries are linear, and SR-Gaussian performs slightly worse than SR-Linear due to the flexibility (i.e., ``over-parameterized'' for linear decision rules) of the Gaussian RBF kernel. In contrast, when true decision boundaries are nonlinear (setting 6), SR-Gaussian outperforms SR-Linear to a large extent as SR-Linear is misspecified in this case. Since we optimize the performance of PLS and D-Learn by modelling the nonlinearity in setting 6, they both enjoy some advantages over the misspecified SR-Linear. In addition, even though our method involves sequential steps and we expect the ``effective" sample size for training each binary classifier may differ when the distribution of the true optimal ITR changes, results from settings 1-4 demonstrate that our method is robust to such variation and performs similarly well across scenarios where proportions of subjects whose $\MD^*(\BX)=1$ are very different. Expectedly, as the sample size increases, misclassification rates decrease, value function estimates get closer to true optimal values (blue dashed lines in Figure S2) and standard deviation estimates of both criteria get smaller. 

\autoref{tab:low_dim_nonparallel} and Figure S3 (supplementary material S.4.3) summarize results corresponding to nonparallel settings 7-10 in the low-dimensional case based on 500 replications. Our proposed SR learning produces the smallest misclassification rates and largest value functions compared to other methods in all settings. Even in the quite complicated case where the nonsmooth boundary between treatment 1 and 2, and the smooth boundary between treatment 2 and 3 are almost ``connected" at the square corner (setting 9), the proposed method still has superior performance and manages to correctly estimate the optimal ITR for more than 90\% of subjects in the testing dataset on average. SR learning also shows the greatest stability in the sense of yielding much smaller standard deviation estimates of misclassification rates and value functions than other methods. 
Interestingly, we observe that ACWL-C1 and ACWL-C2 perform substantially different in settings 8-10 where decision boundaries ``expand" and are nested by nature. Under these scenarios, ACWL-C1 that relies on the minimum contrasts performs much better than ACWL-C2 that uses the maximum contrasts. 


\begin{table}[]
\centering
\caption{Simulation results based on 500 replicates: mean (sd) of misclassification rates and value functions for settings with nonparallel boundaries (7-10) and no noise covariates. The smallest misclassification rate and largest value function for each setting are in bold.}\label{tab:low_dim_nonparallel}
\begin{adjustbox}{width=0.7\textwidth,center}
\begin{tabular}{clccccccccc}
\toprule[0.25ex]
 \rule{0pt}{12pt}
          &  & & \multicolumn{3}{c}{$n=400$} && \multicolumn{3}{r}{$n=800$}                                                        \\ \cline{4-6}  \cline{9-11} \rule{0pt}{14pt}
 Setting        & Method        & & Misclassification &  & Value &&& Misclassification & & Value   \\ \hline
 \rule{0pt}{12pt}
\multirow{6}{*}{7} 
                  & D-Learn             &&  0.14 (0.05)  && 5.12 (0.21)&&& 0.11 (0.03)&& 5.23 (0.12) \\
                  & PLS                 &&  0.14 (0.03)  && 5.11 (0.10)&&& 0.14 (0.02)&& 5.13 (0.07)\\
                  & ACWL-C1         &&  0.16 (0.04)  && 4.92 (0.21)&&& 0.13 (0.03)&& 5.11 (0.16)\\
                  & ACWL-C2         &&  0.14 (0.04)  && 5.09 (0.15)&&& 0.12 (0.03)&& 5.19 (0.11)\\
                  & SR-Gaussian    &&  \textbf{0.03 (0.01)} && \textbf{5.51 (0.05)}&&& \textbf{0.02 (0.01)}&& \textbf{5.56 (0.03)}\\ \hline
 \rule{0pt}{12pt}
 \multirow{6}{*}{8} 
			 & D-Learn     &&  0.25 (0.07)  && 4.64 (0.30)&&& 0.24 (0.06)&& 4.70 (0.24)\\
                  & PLS      &&  0.30 (0.05)  && 4.44 (0.22)&&& 0.29 (0.04)&& 4.48 (0.17)\\
                  & ACWL-C1    && 0.24 (0.04)  && 4.53 (0.28)&&& 0.21 (0.03)&& 4.68 (0.24)\\
                  & ACWL-C2    && 0.52 (0.07)  && 3.52 (0.32)&&& 0.61 (0.07)&& 3.19 (0.28)\\
                  & SR-Gaussian    &&  \textbf{0.05 (0.01)}  && \textbf{5.47 (0.06)}&&& \textbf{0.03 (0.01)}&& \textbf{5.55 (0.03)}\\ \hline
 \rule{0pt}{12pt}
 \multirow{6}{*}{9} 
			 & D-Learn     && 0.26 (0.08)  && 4.64 (0.34)&&& 0.24 (0.07)&& 4.73 (0.28)\\
                  & PLS      && 0.33 (0.06)  && 4.37 (0.24)&&& 0.31 (0.04)&& 4.46 (0.14)\\
                  & ACWL-C1    && 0.22 (0.05)  && 4.60 (0.29)&&& 0.18 (0.03)&& 4.81 (0.26)\\
                  & ACWL-C2    && 0.57 (0.07)  && 3.34 (0.29)&&& 0.69 (0.07)&& 2.93 (0.28)\\
			 & SR-Gaussian    && \textbf{0.08 (0.01)}  && \textbf{5.35 (0.05)}&&& \textbf{0.06 (0.01)}&& \textbf{5.40 (0.03)}\\ \hline
 \rule{0pt}{12pt}
 \multirow{6}{*}{10} 
			 & D-Learn     && 0.40 (0.09)  && 3.83 (0.58)&&& 0.32 (0.08)&& 4.28 (0.45)\\
                  & PLS      && 0.26 (0.07)  && 4.63 (0.28)&&& 0.23 (0.05)&& 4.76 (0.20)\\
                  & ACWL-C1    && 0.44 (0.04)  && 3.30 (0.37)&&& 0.43 (0.03)&& 3.18 (0.34)\\
                  & ACWL-C2    && 0.70 (0.05)  && 0.90 (0.74)&&& 0.77 (0.03)&& 0.09 (0.50)\\
                  & SR-Gaussian    && \textbf{0.09 (0.02)}  && \textbf{5.25 (0.09)}&&& \textbf{0.05 (0.01)}&& \textbf{5.40 (0.06)}\\ 
\bottomrule[0.25ex]
\end{tabular}%
\end{adjustbox}
\end{table}
We measure classification performance through the misclassification rate, which treats all misclassified cases equally. Comments on the metric that measures the degree of disagreement between $\widehat{\mathcal{D}^*}(\BX)$ and ${\mathcal{D}^*}(\BX)$ are made in the supplementary material Section S.4.4.


\subsubsection{Moderate-dimensional $\mathbf{X}$ (With Noise Covariates)}
In this section, we present simulation results for the case where there are many noise covariates that are irrelevant to estimating the optimal ITR. SR-Select results are obtained by applying variable selection methods proposed in Section \ref{subsec:vs_linear} and Section \ref{subsec:vs_nonlinear} for linear and nonlinear decision boundaries (Gaussian kernel), respectively.

\autoref{tab:high_dim_parallel} and Figure S4 (supplementary material S.4.3) show simulation results of settings 1-6 in the presence of noise covariates ($p=50$) for $n=400$ and 800. For reference, we also present the average (across 500 replicates) of misclassification rates and value functions corresponding to the ``oracle" case of the proposed method (red solid lines in Figure S4), where we know exactly which covariates are true signals in decision functions and exclude all noise covariates when applying the method.


\begin{table}[!htb]
\centering
\caption{Simulation results based on 500 replicates: mean (sd) of misclassification rates and value functions for parallel settings (1-6) with noise covariates ($p=50$). The smallest misclassification rate and largest value function for each setting are in bold.}\label{tab:high_dim_parallel}
\begin{adjustbox}{width=0.75\textwidth,center}
\begin{tabular}{clccccccccc}
\toprule[0.25ex]
 \rule{0pt}{12pt}
          &  & & \multicolumn{3}{c}{$n=400$} && \multicolumn{3}{r}{$n=800$}                                                        \\ \cline{4-6}  \cline{9-11} \rule{0pt}{14pt}
 Setting        & Method        & & Misclassification & & Value &&& Misclassification & & Value   \\ \hline
 \rule{0pt}{12pt}
\multirow{6}{*}{1} 
			 & D-Learn             &&  0.25 (0.05) 
  & & 4.00 (0.19)&&& 0.19 (0.04)&& 4.27 (0.15)\\
                  & PLS                 &&  0.23	(0.03)
  & & 4.07 (0.11)&&& 0.16 (0.02)&& 4.37 (0.09)\\
                  & ACWL-C1         &&  0.38 (0.03)
  & & 3.42 (0.16)&&& 0.31 (0.02)&& 3.76 (0.10)\\
                  & ACWL-C2         &&  0.39 (0.03)
  & & 3.41 (0.15)&&& 0.31 (0.02)&& 3.76 (0.09)\\
                  & SR-Linear         && 0.28 (0.03)
  & & 3.86 (0.13)&&& 0.16 (0.03)&& 4.36 (0.10)\\
                  & SR-Linear-Select    &&  \textbf{0.18 (0.03)}
  & & \textbf{4.30 (0.12)}&&& \textbf{0.10 (0.02)}&& \textbf{4.61 (0.08)}\\ \hline
 \rule{0pt}{12pt}
 \multirow{6}{*}{2} 
			      & D-Learn     &&  0.23 (0.03) & & 4.05 (0.12)&&& 0.19 (0.02)&& 4.20 (0.09)\\
                  & PLS      &&  0.18 (0.02)  & & 4.24 (0.07)&&& 0.16 (0.01)&& 4.32 (0.05)\\
                  & ACWL-C1    && 0.31 (0.03)  & & 3.64 (0.14)&&& 0.26 (0.02)&& 3.88 (0.09)\\
                  & ACWL-C2    && 0.31 (0.03)  &  & 3.63 (0.13)&&& 0.26 (0.02)&& 3.90 (0.08)\\
                  & SR-Linear    &&  0.24 (0.03)  & & 4.01 (0.14)&&& 0.14 (0.02)&& 4.43 (0.07)\\
                  & SR-Linear-Select && \textbf{0.15 (0.03)}  & & \textbf{4.38 (0.13)}&&& \textbf{0.09 (0.02)}&& \textbf{4.62 (0.06)}\\ \hline
 \rule{0pt}{12pt}
 \multirow{6}{*}{3} 
			      & D-Learn     && 0.22 (0.02)  & & 4.15 (0.10)&&& 0.19 (0.02)&& 4.29 (0.09)\\
                  & PLS      && 0.20 (0.03)  & & 4.22 (0.11)&&& 0.17 (0.02)&& 4.39 (0.08)\\
                  & ACWL-C1    && 0.31 (0.02)  & & 3.69 (0.13)&&& 0.26 (0.02)&& 3.94 (0.08)\\
                  & ACWL-C2    && 0.31 (0.03)  & & 3.69 (0.14)&&& 0.26 (0.02)&& 3.96 (0.09)\\
                  & SR-Linear   && 0.24 (0.02)  & & 3.99 (0.09)&&& 0.16 (0.02)&& 4.33 (0.08)\\
                  & SR-Linear-Select    && \textbf{0.18 (0.03)}  && \textbf{4.25 (0.12)}&&& \textbf{0.10 (0.02)}&& \textbf{4.57 (0.09)}\\ \hline
 \rule{0pt}{12pt}
 \multirow{6}{*}{4} 
			      & D-Learn     && 0.22 (0.04)  & & 4.14 (0.15)&&& 0.17 (0.02)&& 4.37 (0.10)\\
                  & PLS      && 0.19 (0.02)  & & \textbf{4.27 (0.10)}&&& 0.14 (0.02)&& 4.48 (0.07)\\
                  & ACWL-C1    && 0.36 (0.02)  & & 3.36 (0.14)&&& 0.31 (0.02)&& 3.63 (0.11)\\
                  & ACWL-C2    && 0.36 (0.03)  & & 3.37 (0.15)&&& 0.30 (0.01)&& 3.64 (0.10)\\
                  & SR-Linear  && 0.25 (0.02)  & & 3.87 (0.13)&&& 0.16 (0.02)&& 4.32 (0.08)\\
                  & SR-Linear-Select    && \textbf{0.17 (0.03)}  & & 4.25 (0.13)&&& \textbf{0.11 (0.02)}&& \textbf{4.52 (0.07)}\\ \hline
 \rule{0pt}{12pt}
 \multirow{6}{*}{5} 
			      & D-Learn     && \textbf{0.35 (0.06)}  & & \textbf{3.61 (0.24)} &&& 0.28 (0.04)&& 3.91 (0.17)\\
                  & PLS      && 0.39 (0.04)  &  & 3.41 (0.16)&&& 0.31 (0.03)&& 3.78 (0.13)\\
                  & ACWL-C1    && 0.54 (0.03)  & & 2.31 (0.23)&&& 0.48 (0.03)&& 2.74 (0.17)\\
                  & ACWL-C2    && 0.53 (0.03)  &  & 2.45 (0.19)&&& 0.45 (0.02)&& 2.89 (0.13)\\
                  & SR-Linear    && 0.43 (0.02)  &  & 2.92 (0.18)&&& 0.31 (0.03)&& 3.63 (0.14)\\
                  & SR-Linear-Select    && \textbf{0.35 (0.04)}  & & 3.45 (0.22)&&& \textbf{0.21 (0.03)}&& \textbf{4.09 (0.14)}\\ \hline
 \rule{0pt}{12pt}
 \multirow{6}{*}{6} 
			      & D-Learn     && 0.39 (0.08)  & & 4.83 (0.42)&&& 0.32 (0.07)&& 5.18 (0.33)\\
                  & PLS      && 0.34 (0.03)  & & 5.14 (0.13)&&& 0.26 (0.01)&& 5.44 (0.06)\\
                  & ACWL-C1    && 0.52 (0.05)  & & 3.71 (0.47)&&& 0.43 (0.03)&& 4.48 (0.21)\\
                  & ACWL-C2    && 0.54 (0.04)  & & 3.59 (0.39)&&& 0.46 (0.04)&& 4.39 (0.26)\\
                  & SR-Gaussian    && 0.54 (0.02)  & & 3.68 (0.39)&&& 0.52 (0.02)&& 3.91 (0.24)\\
                  & SR-Gaussian-Select    && \textbf{0.27 (0.06)}  & & \textbf{5.38 (0.30)}&&& \textbf{0.14 (0.04)}&& \textbf{5.96 (0.15)}\\ 
\bottomrule[0.25ex]
\end{tabular}%
\end{adjustbox}
\end{table}

Our proposed method with variable selection (SR-Select) has competitive performance in settings with noise covariates. As expected, misclassification rates and value functions get closer to those from the ``oracle" case as sample size gets larger. When the training dataset size is 400, SR-Select only has marginal gains over PLS and D-Learn in general (and yields slightly worse value functions than D-Learn in setting 5). When training sample size increases to 800, the advantage of SR-Select becomes more pronounced. Since both D-Learn and PLS perform variable selection while estimating the optimal ITR, they do better than SR learning without variable selection in most settings, especially when $n=400$. However, under the large sample size scenario ($n=800$) for linear settings (1-5), SR learning without variable selection performs similarly well as these two methods which incorporate the variable selection feature. We note that even though tree-based methods ACWL-C1 and ACWL-C2 contain intrinsic variable selection via the node-splitting process, it seems that they may not be able to capture true decision boundaries, nor to select relevant covariates for estimating the boundaries in these settings. 
In addition, SR-Select results for setting 6 further confirm that SODA appears to be sufficient for covariate screening in nonlinear settings although it only considers first-order and second-order terms: in this setting, true underlying decision boundaries are nonlinear and involve multiple nonlinear functions of covariates besides second-order polynomials, but SODA still helps reduce the misclassification rate substantially, especially when $n=800$, in which situation SR-Select approaches the ``oracle" case. A possible reason is that the Taylor series expansion of a function up to the second-order terms gives sufficiently good approximations in many cases.



\begin{table}[h]
\centering
\caption{Simulation results based on 500 replicates: mean (sd) of misclassification rates and value functions for nonparallel settings (7-10) with noise covariates ($p=50$). The smallest misclassification rate and largest value function for each setting are in bold.}\label{tab:high_dim_nonparallel}
\begin{adjustbox}{width=0.75\textwidth,center}
\begin{tabular}{clccccccccc}
\toprule[0.25ex]
 \rule{0pt}{12pt}
          &  & & \multicolumn{3}{c}{$n=400$} && \multicolumn{3}{r}{$n=800$}                                                        \\ \cline{4-6}  \cline{9-11} \rule{0pt}{14pt}
 Setting       & Method        & & Misclassification & & Value &&& Misclassification && Value   \\ \hline
 \rule{0pt}{12pt}
\multirow{6}{*}{7} 
                  & D-learn             &&  0.14 (0.04)  & & 5.09 (0.16)&&& 0.11 (0.02)&& 5.21 (0.08)\\
                  & PLS                 &&  0.24	(0.03)  &  & 4.73 (0.12)&&& 0.19 (0.02)&& 4.94 (0.07)\\
                  & ACWL-C1         &&  0.24 (0.05)  & & 4.57 (0.25)&&& 0.15 (0.03)&& 4.97 (0.15)
\\
                  & ACWL-C2         &&  0.25 (0.05)  & & 4.62 (0.22)&&& 0.15 (0.03)&& 5.04 (0.12)\\
                  & SR-Gaussian    &&  0.32 (0.03) & & 4.30 (0.17)&&& 0.25 (0.01)&& 4.66 (0.06)\\ 
                  & SR-Gaussian-Select    &&  \textbf{0.08 (0.03)} & & \textbf{5.37 (0.14)}&&& \textbf{0.03 (0.01)}&& \textbf{5.55 (0.05)}\\\hline
 \rule{0pt}{12pt}
 \multirow{6}{*}{8} 
			 & D-Learn     &&  0.48 (0.10)  & & 3.46 (0.73)&&& 0.38 (0.10)&& 4.08 (0.57)\\
                  & PLS      &&  0.36 (0.02)  & & 4.23 (0.10)&&& 0.34 (0.02)&& 4.30 (0.07)\\
                  & ACWL-C1    && 0.45 (0.07)  && 3.10 (0.75)&&& 0.31 (0.04)&& 4.18 (0.24)\\
                  & ACWL-C2    && 0.52 (0.06)  & & 2.58 (0.73)&&& 0.44 (0.04)&& 3.79 (0.20)\\
                  & SR-Gaussian    &&  0.59 (0.14)  && 2.01 (0.54)&&& 0.59 (0.14)&& 2.07 (0.54)\\ 
                  & SR-Gaussian-Select    &&  \textbf{0.06 (0.03)}  & & \textbf{5.39 (0.12)}&&& \textbf{0.03 (0.01)}&& \textbf{5.53 (0.04)}\\ 
                  \hline
 \rule{0pt}{12pt}
 \multirow{6}{*}{9} 
			 & D-Learn     && 0.53 (0.12)    & & 3.18 (0.84)&&& 0.43 (0.11)&& 3.87 (0.62)\\
                  & PLS      && 0.43 (0.02)  & & 3.92 (0.10)&&& 0.42 (0.02)&& 3.96 (0.07)\\
                  & ACWL-C1    && 0.45 (0.06)  & & 3.07 (0.74)&&& 0.30 (0.05)&& 4.21 (0.29)\\
                  & ACWL-C2    && 0.53 (0.07)  & & 2.52 (0.83)&&& 0.47 (0.04)&& 3.64 (0.22)\\
			 & SR-Gaussian    && 0.64 (0.16)   & & 1.69 (0.79)&&& 0.62 (0.17)&& 1.66 (0.79)\\ 
			 & SR-Gaussian-Select    && \textbf{0.11 (0.03)}  & & \textbf{5.20 (0.18)}&&& \textbf{0.07 (0.01)}&& \textbf{5.36 (0.06)}\\ 
			 \hline
 \rule{0pt}{12pt}
 \multirow{6}{*}{10} 
			 & D-Learn     && 0.62 (0.07)  & & 1.60 (1.23)&&& 0.52 (0.09)&& 3.04 (0.88)\\
                  & PLS      && 0.45 (0.03)  &  & 3.80 (0.17)&&& 0.40 (0.02)&& 4.06 (0.09)\\
                  & ACWL-C1    && 0.63 (0.06)  & & -0.86 (2.05)&&& 0.50 (0.03)&& 2.76 (0.39)\\
                  & ACWL-C2    && 0.70 (0.04)  & & -2.40 (1.36)&&& 0.63 (0.04)&& 1.39 (0.85)\\
                  & SR-Gaussian    && 0.71 (0.04)  & & -2.57 (1.82)&&& 0.69 (0.04)&& -2.82 (1.45)\\ 
                  & SR-Gaussian-Select    && \textbf{0.22 (0.09)}  && \textbf{4.61 (0.67)}&&& \textbf{0.07 (0.03)}&& \textbf{5.34 (0.12)}\\ 
\bottomrule[0.25ex]
\end{tabular}%
\end{adjustbox}
\end{table}
In \autoref{tab:high_dim_nonparallel} and Figure S5 (supplementary material S.4.3), we report simulation results corresponding to moderate-dimensional $\mathbf{X}$ ($p=50$) for nonparallel settings. Similar to parallel settings, we also present mean misclassification rates and value functions of the ``oracle" case as a reference (red solid lines in Figure S5). Unsurprisingly, the performance of the proposed method without variable selection (SR-Gaussian) is seriously affected by the presence of noise covariates and is worse than the other methods that inherently select important variables when making treatment decisions. However, the two-stage variable selection method introduced in Section \ref{subsec:vs_nonlinear} (SR-Gaussian-Select) improves SR-Gaussian substantially and achieves much better performance than D-Learn, PLS, ACWL-C1 and ACWL-C2 in all settings: it leads to much smaller misclassification rates and much larger value functions for both $n=400$ and $n=800$. In particular, it gets very close to the ``oracle" case when $n=800$. 


\section{An Application to the INTERVAL Trial} \label{sec:application}
In the longer term, ageing populations will demand more blood transfusions. In parallel, the pool of young donors is decreasing and maintenance of the blood supply may be challenged by difficulties in attracting and retaining the next generation of blood donors \citep{williamson2013challenge,greinacher2016blood,will2019blood}. The National Health Service Blood and Transplant (NHSBT) in England is investigating making blood donation safer and more sustainable by developing personalized approaches to donation and tailoring the inter-donation interval to the capacity of donors to give blood safely \citep{moore2014interval}. Of particular interest is maintaining blood supplies in universal blood groups (e.g. O negative) that can be used in transfusions for any blood type and ensuring the safety of blood donation, in terms of maintaining iron stores, among high risk groups (e.g. young female donors).


We apply our proposed method to the data from 884 female donors in the INTERVAL trial (introduced in Section \ref{sec:intro}) who were younger than 40 and had O negative blood type. In INTERVAL, female donors were randomly assigned (1:1:1) to the 16-week (standard), 14-week, and 12-week inter-donation intervals. These 3 options can be considered as 3 ordinal ``treatments" and the reference arm is the 16-week inter-donation interval since it is the standard clinical practice and also the safest option. The primary finding from the trial was that there was a substantial increase in the amount of blood collected during the 2-year trial period as a result of the increased donation frequency without major adverse effects on overall
quality of life, physical activity, or cognitive function. However, increased donation frequency led to a greater number of deferrals
(temporary suspension of donors from giving blood) due to low haemoglobin (Hb) \citep{angelantonio2017interval}. We consider the total units of blood (a full donation unit contains 470 ml of whole blood \citep{JPAC2019}) collected by the blood service per donor over the 2-year trial period as the benefit outcome (denoted by $G$), and the number of deferrals for low Hb per donor during the same period as the risk outcome (denoted by $R$). When recommending the optimal inter-donation interval to a blood donor, we should account for the trade-off between the benefit and the risk. Therefore, we construct a utility outcome which ``discounts'' the total units of blood collected by the increased incidences of low Hb deferrals as $U=G-b\times R$, where $b$ is the trade-off parameter reflecting the equivalent benefit loss for one unit increase in the risk, and we seek the individualized donation strategy that maximizes the expected value of the utility score $U$. We note that the trade-off parameter $b$ should be specified based on clinicians' domain knowledge. In the following analysis, we examine two values for $b$, namely $b=2$ and $b=3$, both of which are considered reasonable for the donor subgroup we study (young female donors with universal blood type) by our medical colleagues in the blood service as these values reflect the potential costs of low Hb deferrals incurred by reduced efficiency of collection and reduced donor retention \citep{hillgrovedonorloss2011,will2019blood}. For example, $b=2$ implies that for one extra low Hb deferral per donor attendances, the equivalent loss in the amount of blood collected by the blood service per donor (due to, for example, potential loss of donors following deferrals) is 2 units (over 2 years). 
We consider 14 covariates measured at baseline for estimating the optimal ITR, including age, body mass index (BMI), Short Form Health Survey version 2 (SF-36v2) summary scores (physical component score, mental component score), new or returning donor status, 2-year donation history (i.e., units of whole blood donations in the 2 years before enrollment into the trial), six routine blood measurements (white blood cell count, red blood cell count, haemoglobin level, platelet count, mean corpuscular haemoglobin, mean corpuscular volume), and two blood-based biomarkers (ferritin, transferrin). 


Similar to simulation studies, we compare the proposed SR learning (SR and SR-Select) with D-learn, PLS, ACWL-C1, and ACWL-C2, and all tuning parameters are chosen by 5-fold cross-validation. We focus on linear decision rules for ease of interpretation. To evaluate the performance of each method, we randomly split the data into five almost equal-sized parts and repeat this procedure 100 times. Within each split, we use four parts as the training dataset to learn the optimal rule and predict the optimal inter-donation interval based on the estimated rule for donors in the remaining part (testing dataset). We calculate proportions of donors assigned to each inter-donation interval option and the empirical value function on the testing set. In addition, we also estimate the ``ITR effect", $\delta$, of the estimated rule on the testing set as follows:
$$\widehat{\delta}(\widehat{\mathcal{D}^*})=\frac{\mathbb{P}_{n_{\text{test}}}[YI(A=\widehat{\mathcal{D}^*}(\BX))/P(A|\BX)]}{\mathbb{P}_{n_{\text{test}}}[I(A=\widehat{\mathcal{D}^*}(\BX))/P(A|\BX)]}-\frac{\mathbb{P}_{n_{\text{test}}}[YI(A \neq \widehat{\mathcal{D}^*}(\BX))/P(A|\BX)]}{\mathbb{P}_{n_{\text{test}}}[I(A\neq\widehat{\mathcal{D}^*}(\BX))/P(A|\BX)]},$$
which measures the difference in the empirical value function between the strategy that assigns donors according to the estimated rule and that assigns donors to inter-donation intervals different from the estimated rule \citep{qiu2018lineartrt}. As discussed in the ``practical guide to support vector classification'' \citep{hsu2010svmguide}, linearly scaling each attribute to the same range (e.g. $[-1,1]$) avoids numerical challenges in the kernel calculation, we therefore linearly scale each covariate in the training dataset to $[-1,1]$ before applying SR learning and use the same method to scale covariates in the testing dataset. 



\begin{figure}[]
\centering
\includegraphics[width=0.95\textwidth]{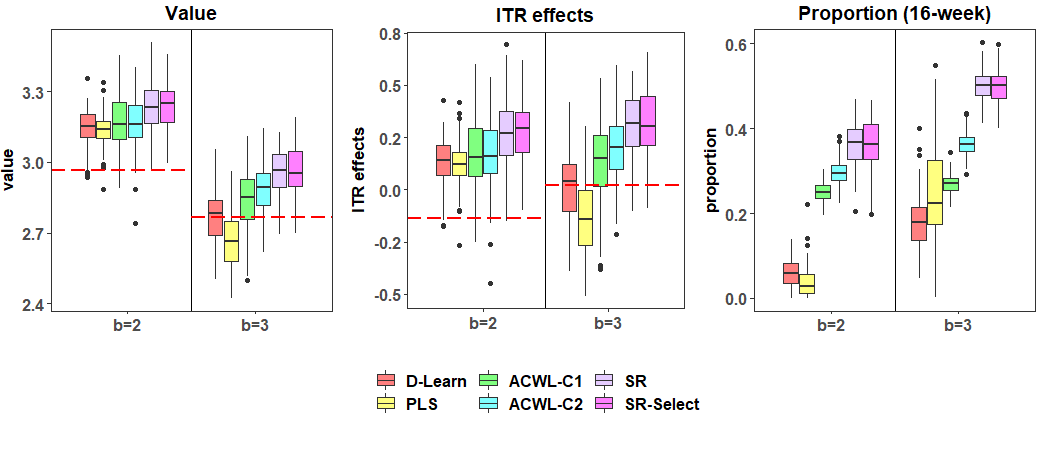}
\caption{Analysis results of the INTERVAL data when the trade-off parameter $b$ in the utility function takes values of 2 and 3: boxplots of estimated value functions, ITR effects, and proportions of donors assigned to the 16-week inter-donation interval (safest option) evaluated on testing datasets based on 100 repetitions of 5-fold cross-validation. Red dashed lines represent value functions/ITR effects under the current clinical practice (assign all female donors to the 16-week inter-donation INTERVAL).}\label{fig:interval_results}
\end{figure}

\autoref{fig:interval_results} plots distributions of estimated value functions, ITR effects and proportions of donors assigned to the 16-week inter-donation interval (safest option) based on 100 repetitions of 5-fold cross-validation, and \autoref{tab:interval_results} summarizes means and standard deviations of cross-validated value functions, ITR effects and donor assignment proportions. For comparison, we also present results corresponding to three non-personalized rules where we assign all donors to the 16-week (current clinical practice), 14-week, or 12-week inter-donation interval, respectively (\autoref{tab:interval_results}).

\begin{table}[!htb]
\centering
\caption{Analysis results of the INTERVAL data when the trade-off parameter $b$ in the utility function takes values of 2 and 3: mean (sd) of donor assignment proportions, value functions, and ITR effects evaluated on testing datasets over 100 repetitions of 5-fold cross-validation for personalized and non-personalized rules. The largest value functions and ITR effects are in bold. Value functions and ITR effects corresponding to the current clinical practice are underlined.}
\label{tab:interval_results}
\centerline{\scalebox{0.72}{
\begin{tabular}{lccccccc}
\toprule[0.25ex]
 \multicolumn{8}{c}{$b=2$}\\
\hline
 \rule{0pt}{14pt}
& \multicolumn{3}{c}{Assignment Proportion (\%)} &  & \multirow{2}{*}{Value} && \multirow{2}{*}{ITR Effects}\\ \cline{2-4}
 \rule{0pt}{14pt}
          & 16 weeks  & 14 weeks  & 12 weeks &  & &&                         \\ \hline
D-Learn & 5.9 (3.4)& 7.6 (4.2) &86.6 (5.3) &  &3.144 (0.081) && 0.131 (0.121) \\ 
PLS     & 3.8  (3.7)& 9.0 (5.5) &87.2 (6.7) &  &3.137 (0.072) && 0.125 (0.110) \\ 
ACWL-C1 & 25.1 (2.1)& 29.4 (2.3) &45.6 (2.4) &  &3.167 (0.111) && 0.168 (0.169) \\ 
ACWL-C2 & 29.6 (3.0)& 27.3 (2.9) &43.1 (3.2) &  &3.167 (0.107) && 0.169 (0.159) \\ 
SR & 36.2 (5.3) & 9.2 (4.5) &54.6 (4.7) &  &3.233 (0.101) &&0.269 (0.154) \\ 
SR-Select & 35.9 (6.3)& 8.8 (5.2) &55.3 (5.5) &  &\textbf{3.236 (0.096)} &&\textbf{0.272 (0.145)} \\ \midrule
Fixed-16 & 100.0 (0.0) & 0.0 (0.0) &0.0 (0.0) &  &\underline{2.964 (0.014)} &&\underline{-0.135 (0.015)} \\ 
Fixed-14 & 0.0 (0.0) & 100.0 (0.0) &0.0 (0.0) &  &3.028 (0.014) &&-0.041 (0.015) \\ 
Fixed-12 & 0.0 (0.0) & 0.0 (0.0) &100.0 (0.0) &  &3.162 (0.015) &&0.166 (0.017) \\ 
\bottomrule[0.25ex]
\toprule[0.25ex]
 \multicolumn{8}{c}{$b=3$}\\
\hline
 \rule{0pt}{14pt}
& \multicolumn{3}{c}{Assignment Proportion (\%)} &  &  \multirow{2}{*}{Value} && \multirow{2}{*}{ITR Effects}\\ \cline{2-4}
 \rule{0pt}{14pt}
          & 16 weeks  & 14 weeks  & 12 weeks &  & &&                         \\ \hline
D-Learn & 18.0 (6.0)& 13.3 (5.4) &68.7 (7.5) &  &2.770 (0.114) && 0.020 (0.169) \\ 
PLS     & 24.4 (11.1)& 9.5 (8.1) &66.1 (11.3) &  &2.667 (0.114) && -0.136 (0.172) \\ 
ACWL-C1 & 27.0 (2.3)& 28.1 (2.3) &44.9 (2.7) &  &2.843 (0.136) && 0.134 (0.206) \\ 
ACWL-C2 & 36.3 (3.1)& 25.4 (3.1) &38.3 (3.3) &  &2.890 (0.111) && 0.205 (0.167) \\ 
SR & 50.1 (3.4) & 8.5 (4.1) &41.4 (3.8) &  &2.957 (0.095) && 0.307 (0.144) \\ 
SR-Select & 49.9 (3.6)& 7.5 (3.8) &42.6 (4.1) &  &\textbf{2.966 (0.101)} && \textbf{0.321 (0.152)} \\ \midrule
Fixed-16 & 100.0 (0.0) & 0.0 (0.0) &0.0 (0.0) &  &\underline{2.768 (0.014)} &&\underline{0.020 (0.016)} \\ 
Fixed-14 & 0.0 (0.0) & 100.0 (0.0) &0.0 (0.0) &  &2.752 (0.016) &&-0.003 (0.017) \\ 
Fixed-12 & 0.0 (0.0) & 0.0 (0.0) &100.0 (0.0) &  &2.745 (0.016)  &&-0.016 (0.016) \\ \bottomrule[0.25ex]
\end{tabular}
}
}
\end{table}

We observe that for both $b=2$ and $b=3$, SR and SR-Select outperform other ITR estimation methods by achieving larger values and larger ITR effects on the utility outcome. In addition, by comparing donor assignment proportions across personalized rules, we find that ITRs estimated by our proposed methods are less ``aggressive" than others in the sense that both SR and SR-Select assign more donors to the safest and currently-implemented inter-donation interval (16-week), especially when $b=3$, in which case low Hb deferral is considered to incur a larger equivalent loss in blood collection. This feature is highly desirable for the donor subgroup under investigation (young female donors with O negative blood type) since those donors are more vulnerable to iron deficiencies and low Hb levels. 
If we follow individualized donation strategies suggested by our proposed methods and assign more donors to the longest and safest inter-donation interval, we would anticipate a reduction in the average risk (low Hb deferral) compared to using ITRs estimated by other methods.  

We also compare personalized donation strategies resulting from SR/SR-Select with the current clinical practice (Fixed-16). Comparisons in terms of donor assignment proportions suggest that our proposed methods encourage more than half of the donors to donate more frequently than the current practice in order to achieve a larger overall utility score which accounts for both the benefit and the risk.  ITR effects associated with proposed methods are much larger than those of Fixed-16. When $b=2$, the empirical value of SR-Select and Fixed-16 are 3.236 units and 2.964 units, respectively, suggesting that the personalized donation strategy estimated by SR-Select leads to an average increase of 127.84 ml (0.272 units) blood collected (``discounted'' by the low Hb deferral) by the blood service \textit{per donor} over 2 years compared to the current practice. The benefit of SR-Select when $b=3$ can be calculated in a similar way, which yields 93.06 ml (0.198 units) additional blood collected (``discounted'' by the low Hb deferral) \textit{per donor} over 2 years. It is likely that estimated ITRs based on INTERVAL data are generalizable to the general UK blood donor pool, since \citet{Moore2016} showed that INTERVAL trial participants were broadly representative of the national donor population of England. According to the NHSBT blood donation database, there are about 23600 female donors under 40 with O negative blood group in the UK general donor population, implying an increase of approximately 3000 liters of blood collected from this donor subgroup when $b=2$ (and roughly 2200 liters when $b=3$) in a 2-year period by implementing this personalized donation strategy. 


To assess effects of different covariates on decision boundaries, we examine the coefficients of linear decision rules estimated by applying the proposed methods to data on all 884 donors using scaled baseline covariates (details described in the supplementary material S.5). We calculate the ``standardized'' absolute effects as a measure of the ``covariate importance'' in linear decision rules. For both $b=2$ and $b=3$, ITRs estimated by SR suggest that the top two baseline covariates with the largest ``standardized'' absolute effects in determining whether or not a young female donor with O negative blood group can be assigned to a shorter inter-donation interval than the 16-week one are mean corpuscular volume and ferritin (Table S5 in the supplementary material).

\section{Discussion}\label{sec:discussion}
This paper presents a sequential re-estimation learning approach to estimate the optimal ITR among ordinal treatments. By exploiting information on treatment orderings, we decompose the ordinal treatment prediction problem into two sets (sequential and re-estimation) of binary treatment selection subproblems that can be solved using existing methods designed for situations with two treatment options. In particular, we solve each subproblem via a weighted SVM \citep{liu2018aol}, which is computationally efficient and guarantees optimal solutions given the convexity of the underlying optimization problem \citep{boyd2004convexoptimization}. Multiple binary decisions can then be aggregated based on a decision tree that again takes into consideration the ordering information on treatments. The proposed SR learning method applies to both linear and nonlinear decision functions, and empirical results demonstrate that it significantly improves classification accuracy and value functions on unseen data compared to methods that do not account for ordinality of treatments. We note that despite assuming treatment 1 to be the reference arm throughout the paper, our method is robust to whether the least ``intensive" treatment (treatment 1) or the most ``intensive" treatment (treatment $K$) is the reference. In the case where $K$ is regarded as the reference treatment according to clinical knowledge, and interest lies in investigating whether treatments that are less ``intensive" than $K$ should be administered, SR learning can be used in a similar manner by reversing the order of comparisons. For example, for $K=3$, three binary subproblems are \{3\} vs. \{2,1\}, \{3\} vs. \{2\}, and \{2\} vs. \{1\}, respectively.

We also develop variable selection methods for ITR estimation under the proposed framework in order to identify covariates that inform treatment decisions and mitigate ``contamination" from noise covariates on decision-making. We propose an ``embedded" variable selection method for linear decision functions and a two-stage ``wrapper" method for nonlinear decision functions \citep{guyon2003vs}. We note that a two-stage procedure similar to the one used for the nonlinear case can be applied to select covariates and estimate optimal ITRs in the linear case by excluding second-order terms from SODA in the first stage and replacing the Gaussian kernel with linear kernel in the second stage. Much like the justification we provide for nonlinear boundaries, this two-stage method should work well in the linear case too. 

Although we focus on randomized clinical trials in introducing the statistical framework and designing the simulation studies, SR learning is applicable to observational studies if we assume both the consistency and the NUC (typically unverifiable in observational studies) assumptions hold true and the propensity score model is correctly specified.

Several extensions can be explored in future research. We adopt $l_1$ regularization for variable selection in this paper. The $l_1$ penalty can be replaced by other types of penalties. For example, an alternative option is the elastic net penalty, which enjoys some nice properties (i.e., highly-correlated covariates are kept or removed together and the number of selected covariates is not upper-bounded by the sample size), and is especially appropriate for the ``$p\gg n$'' problem and the situation where covariates are highly-correlated \citep{zou2005elasticnet,wang2008svmelasticnet}. 
When applying SR learning to the INTERVAL data, we consider the benefit outcome in conjunction with the risk outcome by constructing a utility score and eliciting the trade-off parameter based on clinician's domain knowledge. \citet{wang2017riskbenefit} addressed a similar problem. They proposed to identify the optimal ITR that maximizes the benefit and controls the average risk under a threshold by solving a constrained optimization problem for applications with two treatment arms. They also discussed potential extensions to handle multiple nominal treatments. It would be interesting to investigate how SR learning for ordinal treatments can be modified to fit into their proposed framework to avoid the construction of a utility score. In addition, we develop SR learning under a single-stage set-up. A natural extension is to generalize SR learning to handle multi-stage decision problems and estimate the optimal dynamic treatment regime (DTR) that maximizes the expected long-term outcome in the ordinal treatment setting \citep{liu2018aol}.


\addtolength{\textheight}{.5in}%

\addtolength{\textheight}{-.3in}%

\section*{Supplementary Materials}


Supplementary materials mentioned in Sections 1-4 are provided online. MATLAB and R codes for implementing the proposed SR learning and reproducing all results presented in this paper are available at \url{https://github.com/yx299/SR}.

\section*{Acknowledgements}
This work was supported by the UK Medical Research Council programme MC\_UU\_00002/2 and the Cambridge International Scholarship provided by the Cambridge Commonwealth, European and International Trust. Participants in the INTERVAL trial were recruited with the active collaboration of NHS Blood and Transplant England (\url{www.nhsbt.nhs.uk}), which has supported field work and other elements of the trial. The academic coordinating centre for INTERVAL was supported by core funding from: NIHR Blood and Transplant Research Unit in Donor Health and Genomics (NIHR BTRU-2014-10024), UK Medical Research Council (MR/L003120/1), British Heart Foundation (SP/09/002; RG/13/13/30194; RG/18/13/33946) and the NIHR [Cambridge Biomedical Research Centre at the Cambridge University Hospitals NHS Foundation Trust]. A complete list of the investigators and contributors to the INTERVAL trial is provided in \citet{angelantonio2017interval}. The academic coordinating centre would like to thank blood donor centre staffs and blood donors for participating in the INTERVAL trial. This work was also supported by Health Data Research UK, which is funded by the UK Medical Research Council, Engineering and Physical Sciences Research Council, Economic and Social Research Council, Department of Health and Social Care (England), Chief Scientist Office of the Scottish Government Health and Social Care Directorates, Health and Social Care Research and Development Division (Welsh Government), Public Health Agency (Northern Ireland), British Heart Foundation and Wellcome. The views expressed in this paper are those of the authors and not necessarily those of the NHS, the NIHR or the Department of Health and Social Care.

\bibliographystyle{apalike}

\bibliography{ordinal_arm}

\end{document}